\documentclass[aip, amsmath, amssymb, reprint]{revtex4-2}

\usepackage{graphicx}
\usepackage{blindtext}
\usepackage{color}
\usepackage{dcolumn}
\usepackage{xcolor}
\usepackage{siunitx}

\begin{document}

\sisetup{per-mode = symbol}

\title{Magnetic microcalorimeter with paramagnetic temperature sensors and integrated dc-SQUID readout for high-resolution X-ray emission spectroscopy}

\author{Matthäus Krantz}
\affiliation{Kirchhoff-Institute for Physics, Heidelberg University, Im Neuenheimer Feld 227, 69120 Heidelberg, Germany.}
\author{Francesco Toschi}
\affiliation{Institute for Astroparticle Physics (IAP), Karlsruhe Institute of Technology (KIT), Hermann-von-Helmholtz-Platz 1, 76344 Eggenstein-Leopoldshafen, Germany.}
\affiliation{Kirchhoff-Institute for Physics (KIP), Heidelberg University, Im Neuenheimer Feld 227, 69120 Heidelberg, Germany.}
\author{Benedikt Maier}
\affiliation{Institute of Experimental Particle Physics (ETP), Karlsruhe Institute of Technology (KIT),
Wolfgang-Gaede-Str. 1, 76131 Karlsruhe, Germany.}
\author{Greta Heine}
\affiliation{Institute of Experimental Particle Physics (ETP), Karlsruhe Institute of Technology (KIT),
Wolfgang-Gaede-Str. 1, 76131 Karlsruhe, Germany.}
\author{Christian~Enss}
\affiliation{Kirchhoff-Institute for Physics (KIP), Heidelberg University, Im Neuenheimer Feld 227, 69120 Heidelberg, Germany.}
\affiliation{Institute for Data Processing and Electronics (IPE), Karlsruhe Institute of Technology (KIT), Hermann-von-Helmholtz-Platz 1, 76344 Eggenstein-Leopoldshafen, Germany.}
\author{Sebastian~Kempf}
\email[]{sebastian.kempf@kit.edu}
\affiliation{Institute of Micro- and Nanoelectronic Systems, Karlsruhe Institute of Technology, Hertzstraße 16, 76187 Karlsruhe, Germany.}
\affiliation{Kirchhoff-Institute for Physics (KIP), Heidelberg University, Im Neuenheimer Feld 227, 69120 Heidelberg, Germany.}

\date{\today}

\begin{abstract}
We present two variants of a magnetic microcalorimeter with paramagnetic temperature sensors and integrated dc-SQUID readout for high-resolution X-ray emission spectroscopy. Each variant employs two overhanging gold absorbers with a sensitive area of \qtyproduct{150 x 150}{\micro\meter} and a thickness of \SI{3}{\micro\meter}, thus providing a quantum efficiency of \SI{98}{\percent} for photons up to \SI{5}{\kilo\electronvolt} and \SI{50}{\percent} for photons up to \SI{10}{\kilo\electronvolt}. The first variant turned out to be fully operational, but, at the same time, to suffer from Joule power dissipation of the Josephson junction shunt resistors, athermal phonon loss, and slew rate limitations of the overall setup. Overall, it only achieved an energy resolution $\Delta E_\mathrm{FWHM} = \SI{8.9}{\electronvolt}$. In the second variant, we introduced an innovative `tetrapod absorber geometry' as well as a membrane-technique for protecting the temperature sensors against the power dissipation of the shunt resistors. By this, the second variant achieves an outstanding energy resolution of $\Delta E_\mathrm{FWHM} = \SI{1.25(18)}{\electronvolt}$ and hence provides, to our knowledge, the present best energy resolving power $E/\Delta E_\mathrm{FWHM}$ among all existing energy-dispersive detectors for soft and tender X-rays.
\end{abstract}

\maketitle

X-ray emission spectroscopy (XES) is an exceptionally powerful tool to study fundamental properties of materials such as chemical states or atomic and electronic structure of constituents \cite{Gro01, Uhl15}. By analyzing the X-ray photons which are emitted by the sample, it provides valuable information that can hardly be obtained by other techniques. However, to exploit its full power, an X-ray detector with outstanding features is required. In this respect, cryogenic microcalorimeters such as superconducting transition-edge sensors (TESs) \cite{Irw05, Ull15, Fri06} or magnetic microcalorimeters (MMCs) \cite{Fle05, Kem18} are a striking detector technology as they combine an excellent energy resolution, a large dynamic range, and a quantum efficiency close to \SI{100}{\percent} in a single device \cite{Dor16, Uhl15, Fri06}. Using an ultra-sensitive thermometer, based on either superconducting (TESs) or paramagnetic (MMCs) materials, as well as an appropriate readout circuit, they convert the energy input into a change of current or magnetic flux, respectively, that can be sensed using superconducting quantum interference devices (SQUIDs). Existing detectors achieved an energy resolution $\Delta E_\mathrm{FWHM}$ of \SI{0.72}{\electronvolt} for \SI{1.5}{\kilo\electronvolt} photons \cite{Lee15} and \SI{1.6}{\electronvolt} for \SI{5.9}{\kilo\electronvolt} photons \cite{Smi12,Kem18}. Nevertheless, they differ severely in size, i.e. the previous record detector, for example, only provides a sensitive area of \qtyproduct{45 x 45}{\micro\meter} \cite{Lee15}.

\begin{figure*}
    \includegraphics[width=1.0\linewidth]{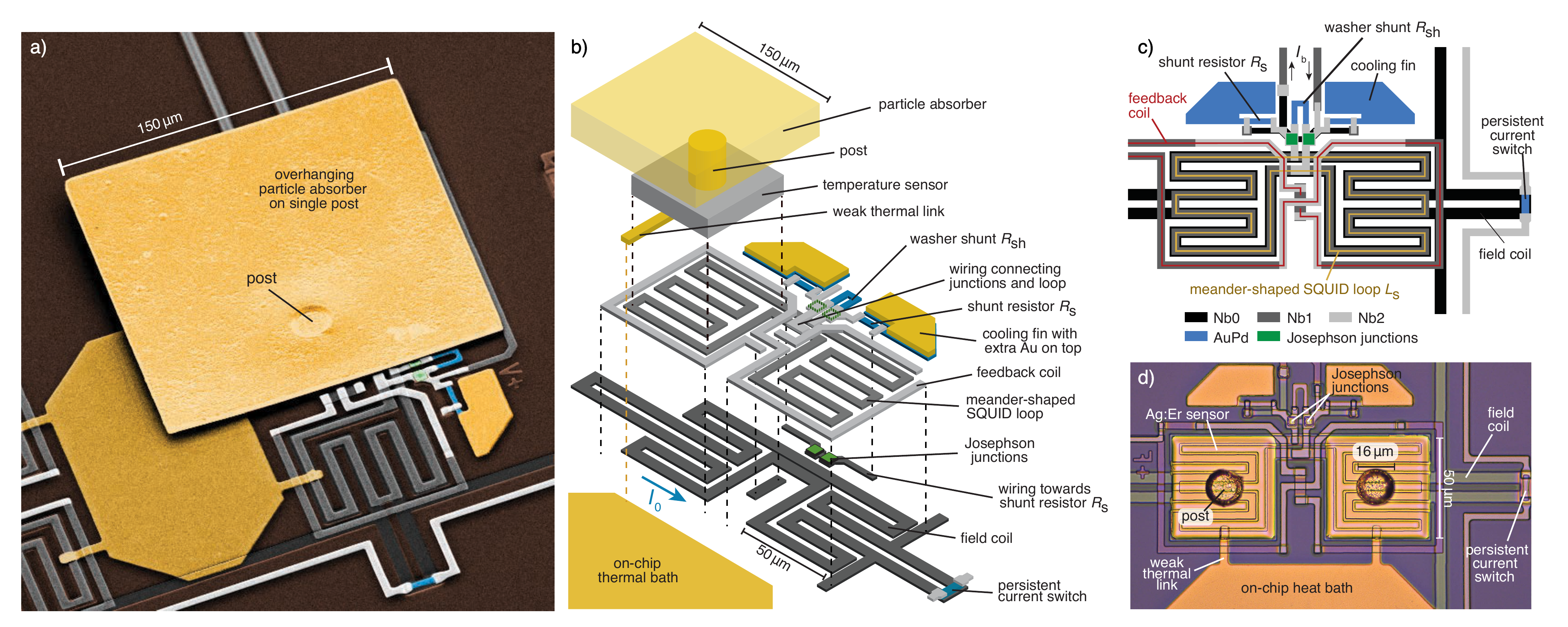}%
    \caption{\label{fig1}
    (a) Colorized SEM picture, (b) exploded-view drawing, (c) simplified layout drawing, and (d) microscope photograph of the first variant of our microcalorimeter with paramagnetic temperature sensors and integrated dc-SQUID readout. For visibility, we show images where we omitted or remover one or two absorbers and/or temperature sensors. For the same reason, the layout drawing in (c) only shows the arrangement of the three Nb layers, the Josephson tunnel junctions as well as the resistive Au:Pd structures.}
\end{figure*}

Among the existing cryogenic detectors, magnetic microcalorimeters stand out in the sense that they simultaneously provide an excellent energy resolution, an intrinsically fast signal rise time, a huge dynamic range, as well as easy calibration and excellent linearity \cite{Kem18}. However, at the same time, they are susceptible to SQUID noise that easily can limit the achievable energy resolution \cite{Kem18}. One possibility to deal with this challenge is to integrate the temperature sensor(s) directly into the SQUID loop \cite{Zak03, Sto05, Kem18}. However, the close vicinity between the resistive junction shunts and the temperature sensors makes such devices rather sensitive to SQUID Joule power dissipation and hampered the usage of such devices in the past. Nonetheless, we revisited this old idea and now present two variants of a magnetic microcalorimeter with paramagnetic temperature sensors and integrated dc-SQUID readout for high-resolution X-ray emission spectroscopy. Each variant employs two overhanging gold absorbers with a sensitive area of \qtyproduct{150 x 150}{\micro\meter} and a thickness of \SI{3}{\micro\meter}, thus providing a quantum efficiency of \SI{98}{\percent} for photons up to \SI{5}{\kilo\electronvolt} and \SI{50}{\percent} for photons up to \SI{10}{\kilo\electronvolt}. While the first variant turned out to be still susceptible to power dissipation by the junction shunts, we successfully included countermeasures in our second variant. By this, the second achieves an outstanding energy resolution of $\Delta E_\mathrm{FWHM} = \SI{1.25(18)}{\electronvolt}$ and hence provides, to our knowledge, the present best energy resolving power $E/\Delta E_\mathrm{FWHM}$ among all existing energy-dispersive detectors for soft and tender X-rays.

\begin{figure*}
    \includegraphics[width=1.0\linewidth]{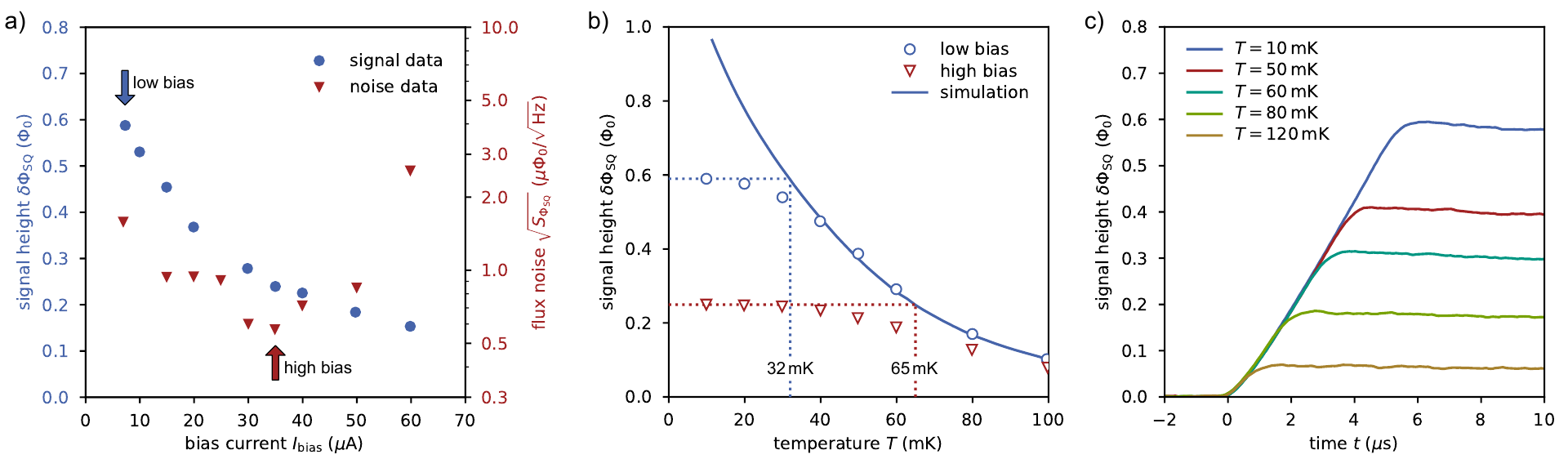}%
    \caption{\label{fig2}
    (a) Signal height $\delta \Phi_\mathrm{SQ}$ and white noise level $\sqrt{S_{\Phi_\mathrm{SQ}}}$ vs. bias current $I_\mathrm{bias}$ used for creating the voltage bias of the detector within a two-stage dc-SQUID configuration. The measurement was performed at based temperature of the cryostat. The arrows indicate the working points referenced in the middle plot.
    (b) Signal height $\delta \Phi_\mathrm{SQ}$ vs. heat bath temperature $T$ when biasing the SQUID with smallest possible bias voltage to achieve the largest detector signals (low bias) as well as when biasing the detector to yield optimum noise performance (high bias). In addition, we show the simulated dependence for the given operation and detector parameters. The deviation between measurement and simulation towards low temperatures is caused by a thermal decoupling of the detector from the heat bath due SQUID Joule heating. The dashed lines are extrapolations from the measured data to the expected values and allow determining the actual detector temperature.    
    (c) Signal height $\delta \Phi_\mathrm{SQ}$ vs. time $t$ for detector signals of different height.
    }
\end{figure*}

Fig.~\ref{fig1} shows a colorized scanning electron microscope (SEM) picture, an exploded-view drawing, a simplified layout drawing as well as a microscope photograph of the first variant of our microcalorimeter. The main SQUID loop is based on two superconducting meander-shaped coils made of Nb that are connected in parallel to the Josephson junctions to form a first-order parallel-gradiometer. The pitch and line width of each coil are $p_2 = \SI{10}{\micro\meter}$ and $w_2 = \SI{4}{\micro\meter}$, respectively. Underneath the SQUID loop, a meander-shaped coil is running essentially replicating the shape of the main SQUID loop. The pitch and line width of this coil are $p_1 = \SI{10}{\micro\meter}$ and $w_1 = \SI{6}{\micro\meter}$, respectively. The line width $w_1$ is chosen slightly larger than that of the SQUID loop to account for the alignment accuracy of our fabrication equipment. This `field coil' is used for generating the bias magnetic field required to magnetize the temperature sensors. For this, a persistent current is injected into the coil using a resistive persistent current switch that is located nearby the SQUID loop. It is worth mentioning that separating field and pickup coil allows simultaneously injecting a persistent current into several microcalorimeters by serially connecting all field coils. In case only one coil would be used, the ground connection of the SQUID forms as parasitic current path hampering a reliable persistent current injection.

The window-type Josephson junctions are made from an Nb/Al-AlO$_x$/Nb trilayer and have a target critical current of $I_\mathrm{c} \approx \SI{6.3}{\micro\ampere}$. Each junction is shunted by a resistor with $R_\mathrm{s} \approx \SI{5}{\ohm}$ that is connected to a cooling fin. To increase its effective volume, the shunt resistor is covered with a sputtered Au layer. To dampen parasitic resonances in the $IV$-characteristic of the SQUID, we connected a resistor with $R_\mathrm{s} = \SI{10}{\ohm}$ in parallel to the SQUID loop. The SQUID loop is inductively coupled to a feedback coil for flux-biasing as well as flux-locked loop operation using a direct-coupled high-speed SQUID electronics.

A Ag:Er temperature sensor with an area of \qtyproduct{50 x 50}{\micro\meter} is placed on top of the meander-shaped SQUID loop. For the first detector variant, the Er concentration and sensor height are \SI{450}{ppm} and \SI{1.2}{\micro\meter}, respectively. Based on the measured detector response and noise characteristics of this first detector, the second detector variant has an Er concentration of \SI{260}{ppm} and a sensor width of \SI{0.8}{\micro\meter}. The overhanging particle absorbers are made of electroplated gold with dimensions of \qtyproduct{150 x 150 x 3}{\micro\meter}, resulting in a heat capacity of each absorber of $C_\mathrm{abs} = \SI{0.1}{\pico\joule\per\kelvin}$ at $T = \SI{20}{\milli\kelvin}$. A single post with a diameter of $d=\SI{16}{\micro\meter}$ connects the absorber with the underlying temperature sensor. The cross-sectional area of the stem is about \SI{8}{\percent} of the total sensor area and is hence rather large as compared to other MMCs \cite{Fle09}. However, we found that for the given absorber geometry (size, thickness) a single post with a rather large diameter is structurally much more stable than several posts with a smaller diameter.

For device characterization, we mounted a prototype device into a $^3$He/$^4$He dilution refrigerator using a custom-made sample holder. The latter was designed to allow the screening of the detector against external magnetic field fluctuations using a superconducting Al shield as well as the operation of the detector in a two-stage dc-SQUID configuration with voltage bias and flux-locked loop. We used an updated version of our home-made 16-dc-SQUID series arrays \cite{Kem17} as low-noise amplifier SQUID as well as a direct-coupled room-temperature SQUID readout electronics \footnote{XXF-1 SQUID Electronics from Magnicon GmbH, Hamburg, Germany.}. We irradiated the detector with X-rays emitted by an $^{55}$Fe calibration source mounted inside the cryostat.

Fig.~\ref{fig2} shows a summary of the most important findings of the characterization of the first detector variant. The detector was fully functional, despite the very high complexity of its manufacture that origins from fourteen different photolithographic layers including three independent Nb layers. We achieved an energy resolution of $\Delta E_\mathrm{FWHM} = \SI{8.9}{\electronvolt}$, the latter being about an order of magnitude worse than our expectation. After a careful investigation/analysis, we found the SQUID Joule heating to be the main source of the performance degradation. When biasing the detector to yield optimum noise performance (high bias), the temperature sensors being located nearby the shunt resistors didn't go below $T \approx \SI{65}{\milli\kelvin}$. In contrast, when biasing the SQUID with smallest possible bias voltage to yield the largest detector signals (low bias), the overall noise level was more than a factor of three higher than under high bias conditions. Moreover, the deviation between the measured and predicted signal size indicated that even for the smallest bias voltage the signal height was about a factor of two smaller than possible. Additionally, we found that the detector suffered slightly from athermal phonon loss due to the rather large fraction between the stem and sensor area as well as strongly from hitting the slew rate limit of the SQUID setup that was determined by the length of the wiring as well as the total SQUID gain. The latter became noticeable by a linear rather than exponential dependence of the time course of the signal rise (see Fig.~\ref{fig2}c).

\begin{figure}
    \includegraphics[width=1.0\linewidth]{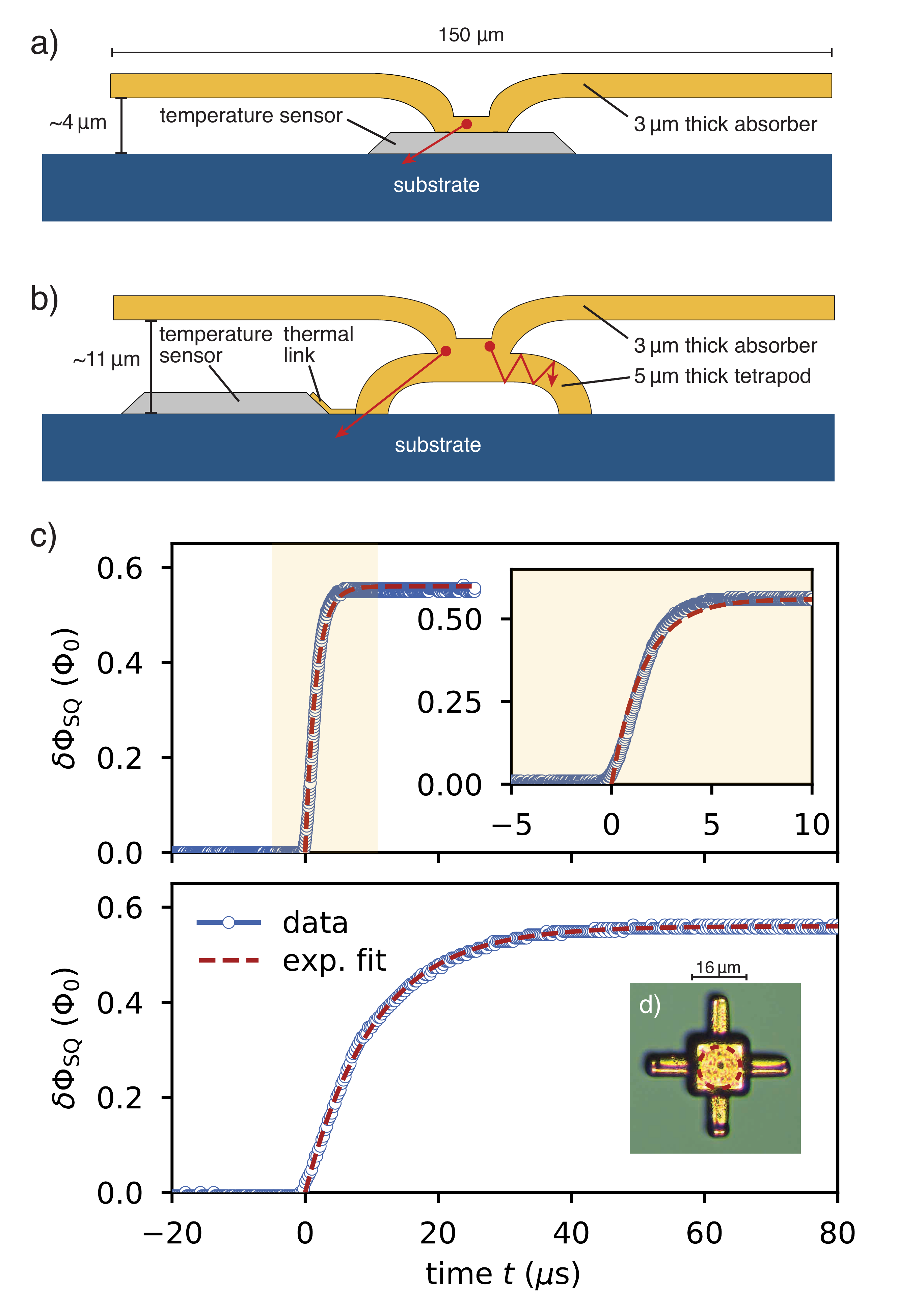}%
    \caption{\label{fig3}
    Illustration of the (a) conventional and (b) tetrapod absorber geometry. The red arrows show possible traveling paths for athermal phonons generated within the absorber. In the conventional geometry, athermal phonons can easily escape to the substrate, while in the tetrapod geometry the cross-sectional area of direct line of sight is significantly reduced. (c) Time course of the signal rise of two prototype detectors of the second variant for which the thermal conductance of the thermal link was varied. The red dashed lines shows exponential fits to the measured detector response. The fast detector (upper plot) is still slew rate limited while the slow detector (lower plot) shows the expected exponential signal rise. For the fast and slow detectors the rise time (time constant of the exponential fit function) is about \SI{2}{\micro\second} and \SI{10}{\micro\second}, respectively. The inset shows a magnification of the marked area. (d) Microscope picture of a finalized tetrapod structure on a test sample. The red circle indicates the area where the metallic post connects absorber and tetrapod.}
\end{figure}

We investigated several methods to reduce the impact of SQUID Joule heating, athermal phonon loss, and slew rate limitation on the detector performance. We successfully resolved the latter two by introducing a `tetrapod absorber geometry' as depicted in Fig.~\ref{fig3}. Here, the particle absorber is not directly attached to the temperature sensor, but instead on top of a four-legged bridge (tetrapod). In this geometry, the direct line of sight between absorber and sensor has a smaller cross-sectional area, significantly reducing the probability for athermal phonons to escape. We placed one tetrapod leg on top of a thermal link made of sputtered gold. By varying the length and width of this link, we can set the signal rise time. We empirically determined that the detector rise is exponential (and hence not slew rate limited) as soon as the rise time is down to about $\sim \SI{10}{\micro\second}$. Though this slow down impacts the time resolution of the detector, the energy resolution is not affected \cite{Ban12} as the effective bandwidth of the detector is smaller.

\begin{figure*}
    \includegraphics[width=1.0\linewidth]{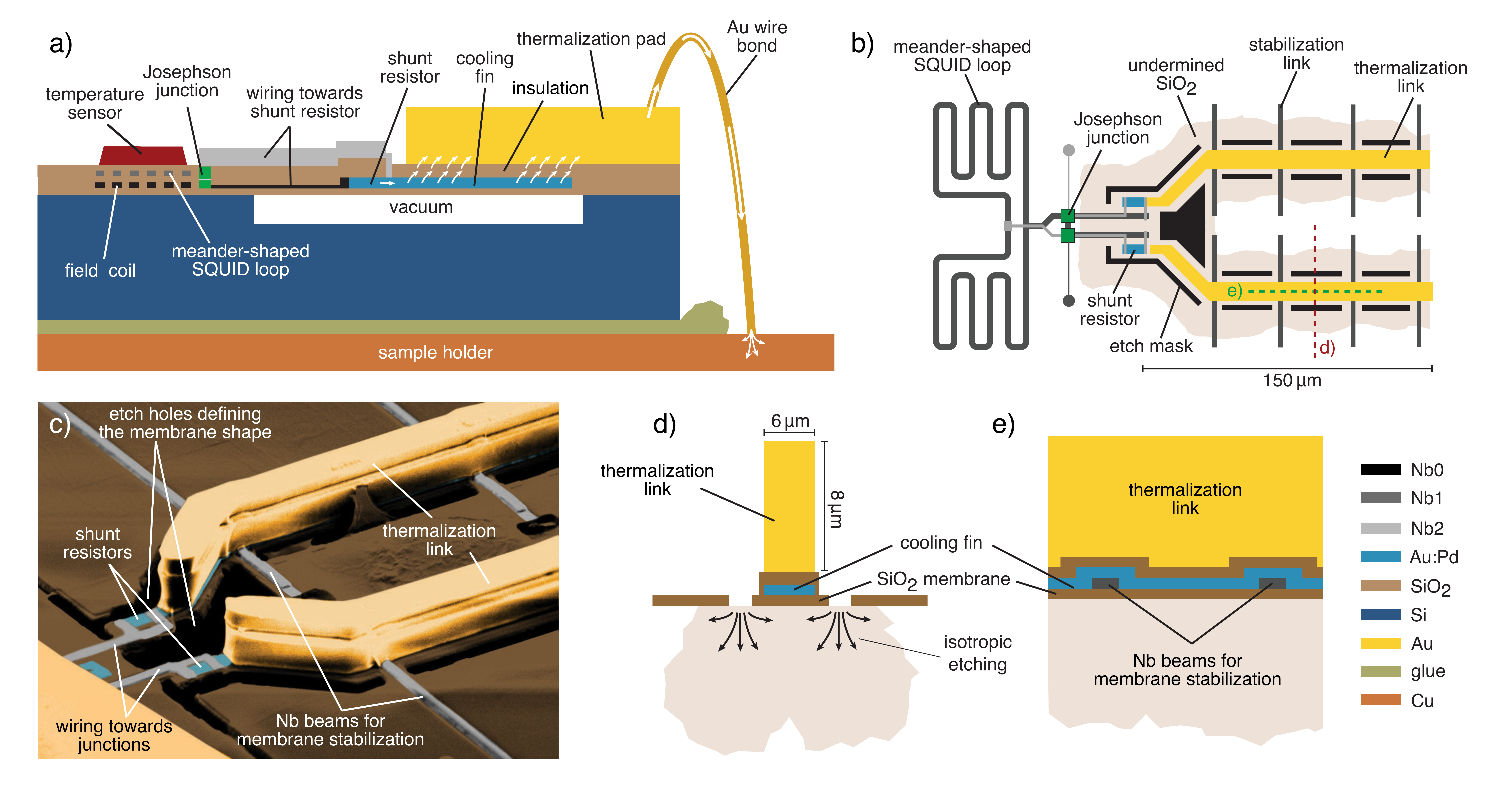}%
    \caption{\label{fig4}
    Schematic illustration of a (a) side view and (b) top view of the membrane arrangement developed to reduce the impact of SQUID Joule heating on the detector performance. (c) Colorized SEM picture of the shunt resistor section of a finalized prototype detector of the second variant. The thermalization bridges as well as stabilization structures are sitting on a SiO$_2$ membrane that is undermined using SF$_6$ based reactive ion etching. (d), (e) Cross-sectional views of the regions marked in b).}
\end{figure*}

\begin{figure*}
    \includegraphics[width=1.0\linewidth]{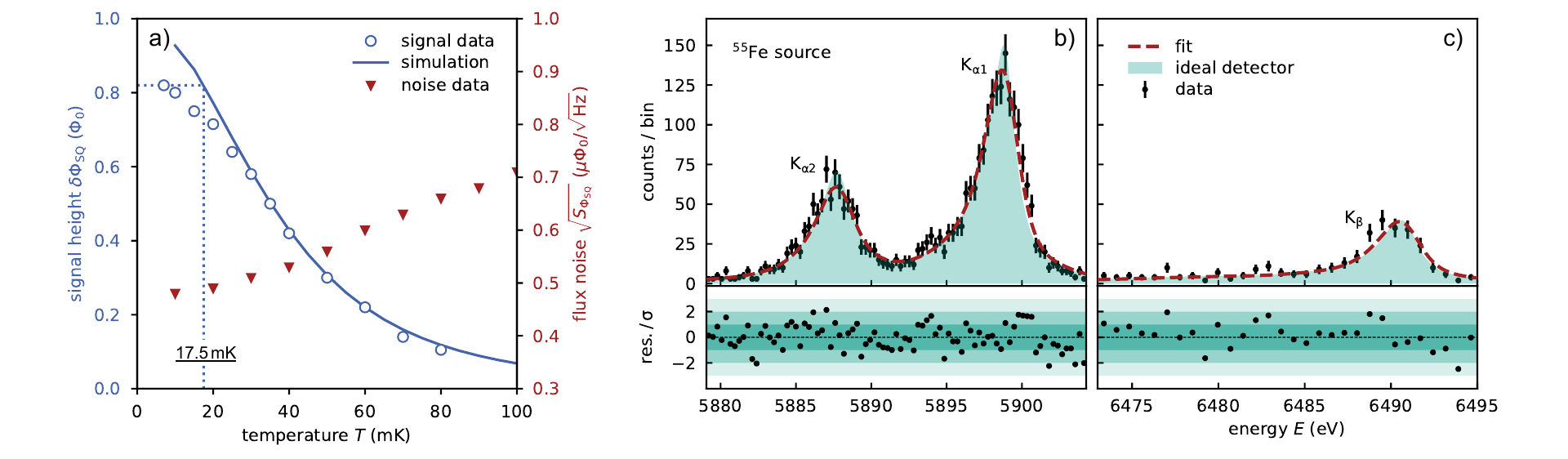}%
    \caption{\label{fig5}
    (a) Signal height $\delta \Phi_\mathrm{SQ}$ and white noise level $\sqrt{S_{\Phi_\mathrm{SQ}}}$ vs. temperature $T$ as well as the expected dependence derived by numerically simulating the detector behavior. The deviation of the measured data from the simulation towards low temperatures is caused by a slight thermal decoupling of the detector from the heat bath due remaining SQUID power dissipation. The dashed line is an extrapolation from the measured data to the expected values and allows determining the actual detector temperature. (b) $K_\alpha$ and (c) $K_\beta$ line of the $^{55}$Fe calibration source used for detector characterization. The solid red lines are the result of a fitting procedure to determine the energy resolution $\Delta E_\mathrm{FWHM} = \SI{1.25(18)}{\electronvolt}$ of the detector. The filled areas illustrate the shape of the expected spectra assuming an ideal detector with $\Delta E_\mathrm{FWHM} = \SI{0}{\electronvolt}$. Comprehensive details about the data analysis and a breakdown of the statistical and systematic errors are discussed by Toschi {\it et al.}\cite{Tos23}.}
\end{figure*}

For reducing the impact of SQUID Joule heating, we investigated several methods, among which a reduction of the shunt resistance as well as backside coating of the detector with a \SI{1}{\micro\meter} thick Au layer. The latter acts as a sink for athermal phonons emitted by the shunts and ballistically traversing through the silicon substrate, and it enhances the thermal coupling between detector and sample holder. In combination, both techniques allowed to lower the actual detector temperature to \SI{25}{\milli\kelvin}, even when using a bias minimizing the SQUID noise. This greatly improved the detector performance and we achieved an energy resolution as low as $\Delta E_\mathrm{FWHM} \simeq \SI{2.4}{\electronvolt}$. Nevertheless, we noticed that the detector performance still remains susceptible to the number of detectors operated within an array. For this reason, we developed a method that is based on placing the shunt resistors on a thin $\mathrm{SiO}_2$ membrane\footnote{We used our standard thermally oxided Si wafers for fabricating the detector. For future fabrications runs, we might investigate whether Si wafers coated with low-stress $\mathrm{Si}_3\mathrm{N}_4$ might simplify the fabrications and enhance the reliabilty and quality.} and to thermally anchor the shunts through a separate metallic link directly to the sample holder. Fig.~\ref{fig4} shows several illustrations of this method and indicates that field coil, SQUID loop as well as both temperature sensors and absorbers are placed on solid substrate. The shunt resistors are still connected to a cooling fin. Below the shunt resistors, the Si substrate is removed by isotropic reactive ion etching. The shunts are hence hovering on a membrane formed by the \SI{240}{\nano\meter} thick thermal $\mathrm{SiO}_2$ electrically insulating the Si wafer. As the $\mathrm{SiO}_2$ membrane is rather fragile, we used Nb beams to stabilize the membrane. Overall, the $\mathrm{SiO}_2$ membrane greatly reduces the thermal conductance between the shunt resistors and the temperature sensors and prohibits that dissipated energy is transferred into the solid substrate. To sink the heat generated by the junction shunts, we provide a separate thermalization pathway to the sample holder by electro-depositing a Au thermalization link on top of the cooling fin as well as on a small part of the solid substrate for structural stability. This thermalization link is coupled to the sample holder via Au wire bonds. Since at MMC operation temperatures the phononic thermal conductance is orders of magnitude lower than the electronic one as well as due to the Kapitza resistance, only a negligible fraction of the dissipated energy is transferred from the Au thermalization link into the Si substrate. Overall, the separation of the shunt resistors from the temperature sensors made it possible to greatly reduce the impact of the SQUID Joule heating on the detector performance and to lower the actual sensor temperature to about \SI{17}{\milli\kelvin} (see Fig.~\ref{fig5}a). We can hence use an optimal SQUID bias to get very low SQUID noise and large signals. Moreover, we noticed that we are only missing about \SI{20}{\percent} of signal size as compared to our detector simulations (see Fig.~\ref{fig5}a). The remaining thermal decoupling of the temperature sensors from the heat bath is likely caused by the washer shunt of the SQUID that is not sitting on the $\mathrm{SiO}_2$ membrane as well as the fact that we potentially loose some heat to the Si substrate via the thermalization links. Both points can be addressed in a future design by placing the washer shunt on the membrane as well as using normal-conducting through-silicon vertical interconnects (vias) for heat sinking of the shunt resistors on the membrane. 

Fig.~\ref{fig5}b) and c) show the energy spectra of the $K_\alpha$ and $K_\beta$ line of the $^{55}$Fe calibration source as acquired with the latest prototype detector of the second detector variant that uses the tetrapod absorber geometry with a signal rise time of about \SI{10}{\micro\meter} as well as a $\mathrm{SiO}_2$ membrane to thermally separate the temperature sensors from the shunt resistors. For the analysis of the acquired signals, we used an optimal filter technique that is described in detail in a separate publication \cite{Tos23}. During analysis, we found that the detector is affected by strong drifts of the heat bath temperature that was not regulated during the measurement as the temperature controller induced severe disturbances to the detector. To (partially) account for the resulting degradation of energy resolution, we performed a temperature correction of the signal heights and excluded the signals with largest temperature deviations. In total, \SI{26}{\percent} of all acquired physical/good signals were discarded. Moreover, we assumed that the detector response is generally Gaussian, but can be potentially affected by athermal phonon loss leading to a low energy tail of the spectra. By performing a sophisticated fitting procedure \cite{Tos23} and minimizing the $\chi^2$ deviation between the measured spectrum and the detector model, we find that athermal phonon loss is negligible and that the energy resolution of the detector is $\Delta E_\mathrm{FWHM} = \SI{1.25(18)}{\electronvolt}$ when excluding the detector signals with largest temperature deviations. When accepting all detector signals (and including athermal phonon loss), the energy resolution is $\Delta E_\mathrm{FWHM} = \SI{1.35(17)}{\electronvolt}$\footnote{We note that the quoted errors contain both, the statistical and systematic errors and refer to Toschi {\it et al.}\cite{Tos23} for a comprehensive breakdown of the different error contributions.}. To our knowledge, the presented detector hence provides the best energy resolution power $E/\Delta E_\mathrm{FWHM}$ of any energy-dispersive detector in the soft and tender X-ray range.

In conclusion, we have presented two variants of a magnetic microcalorimeter with paramagnetic temperature sensors and integrated dc-SQUID readout for high-resolution X-ray emission spectroscopy. Each variant employs two overhanging gold absorbers with a sensitive area of \qtyproduct{150 x 150}{\micro\meter} and a thickness of \SI{3}{\micro\meter}, thus providing a quantum efficiency of \SI{98}{\percent} for photons up to \SI{5}{\kilo\electronvolt} and \SI{50}{\percent} for photons up to \SI{10}{\kilo\electronvolt}. As the first variant suffers from Joule power dissipation of the SQUID, athermal phonon loss and reaching the slew rate limit of the overall setup, we introduced a `tetrapod absorber geometry' as well as a membrane technique for protecting the temperature sensors against the power dissipation of the shunt resistors. By this, the detector achieves an outstanding energy resolution of $\Delta E_\mathrm{FWHM} = \SI{1.25(18)}{\electronvolt}$ and hence provides, to our present knowledge, the best energy resolving power among all existing energy-dispersive X-ray detectors in the energy range up to \SI{10}{\kilo\electronvolt}.

\begin{acknowledgments}
We would like to thank T. Wolf for his support during device fabrication and greatly acknowledge valuable discussions (in alphabetical order) with A. Fleischmann, T. Ferber, M. Klute, and B. v.Krosigk. The research leading to these results has also received funding from the European Union’s Horizon 2020 Research and Innovation Programme, under Grant Agreement No 824109. The research was further supported by the Federal Ministry of Education and Research grant 05P15VHFAA. M. Krantz acknowledges financial support by the HighRR Research Training Group (GRK2058) funded by the German Research Foundation. Moreover, G. Heine and B. Maier received financial support by the Alexander von Humboldt Foundation in the framework of the Alexander von Humboldt Professorship of M. Klute endowed by the Federal Ministry of Education and Research.
\end{acknowledgments}

\section*{Author Declarations}
\subsection*{Conflict of Interest}
The authors have no conflicts to disclose.

\subsection*{Author Contributions}
{\bf Matthäus Krantz:} Conceptualization (equal); Formal Analysis (equal); Investigation (lead); Visualization (lead); Writing – original draft (supporting); Writing/Review \& Editing (equal). {\bf Francesco Toschi:} Formal Analysis (lead); Software (equal); Writing/Review \& Editing (equal). {\bf Benedikt Maier} Formal Analysis (equal); Software (equal); Writing/Review \& Editing (equal). {\bf Greta Heine} Formal Analysis (equal); Software (equal); Writing/Review \& Editing (equal). {\bf Christian Enss:} Conceptualization (supporting); Funding Acquisition (lead); Investigation (equal); Project Administration (supporting); Resources (lead); Supervision (equal); Writing/Review \& Editing (equal). {\bf Sebastian Kempf:} Conceptualization (equal); Formal Analysis (supporting); Investigation (equal); Project Administration (supporting); Resources (equal); Supervision (equal); Visualization (equal); Writing/Original Draft Preparation (lead); Writing/Review \& Editing (equal).

\section*{Data Availability Statement}
The data that support the findings of this study are available from the corresponding author upon reasonable request.

\bibliography{literature}

\end{document}